\documentclass[11pt,a4paper]{article}
\usepackage{graphicx}
\usepackage{hyperref}
\usepackage{comment}
\usepackage{float}

\begin{document}

\title{Web password recovery --- a necessary evil?}
	\author{Fatma Al Maqbali and Chris J Mitchell\\
	Information Security Group, Royal Holloway, University of
	London\\
	\href{mailto:fatmaa.soh@cas.edu.om}{\nolinkurl{fatmaa.soh@cas.edu.om}},
	\href{mailto:me@chrismitchell.net}{\nolinkurl{me@chrismitchell.net}}
}
\date{15th July 2016}
\date{\today} 
\maketitle
\begin{abstract}

Web password recovery, enabling a user who forgets their password to
re-establish a shared secret with a website, is very widely implemented.
However, use of such a fall-back system brings with it additional
vulnerabilities to user authentication. This paper provides a framework within
which such systems can be analysed systematically, and uses this to help gain a
better understanding of how such systems are best implemented. To this end, a
model for web password recovery is given, and existing techniques are
documented and analysed within the context of this model. This leads naturally
to a set of recommendations governing how such systems should be implemented to
maximise security.  A range of issues for further research are also
highlighted.

\end{abstract}

\section{Introduction}

Despite their widely-documented limitations, passwords remain very widely used
for user authentication.  However, although passwords are meant to be
memorized, humans often forget or mislay them.  In some contexts this is easily
managed; for example, in an office environment, a user who forgets their
password for access to a multi-user system can simply see a system
administrator, who authenticates them and issues a new password. However, for
web authentication, the main focus of this paper, it is clearly not so simple;
we refer to the process of re-establishing a password in this case as
\textit{web password recovery}.

Most websites requiring users to create an account support password recovery,
enabling legitimate users who forget their password to continue to use the website. This can be achieved in many ways, including use of a pre-registered
email address or mobile phone number, and/or involving other pre-set fall-back
means of authentication. However, many of these techniques introduce
vulnerabilities which may enable an impostor to falsely change a password,
either causing a denial of service or, in the worst case, enabling the impostor
to authenticate as the user.  In this paper, we analyze these issues by first
introducing a general model for password recovery, and then examining various
existing password recovery options within the context of this model. We also
use the model to look at ways of improving the security and/or usability of
password recovery. 		

The remainder of this paper is structured as follows. In section \ref{Prior
Art} we review prior art on password recovery, noting that much of the existing
work on password recovery does not match the scope of this paper. This is
followed in section \ref{A General Model} by a general model for web password
recovery, intended to capture the commonly used approaches. Section \ref{Model
Components} then reviews the various ways in which password recovery is
performed in practice, within the context of the model. This then enables us in
section \ref{Security and usability issues} to provide a systematic assessment
of the security strengths and weaknesses of existing approaches, along with
potential usability issues. In section \ref{Towards secure password recovery}
we use the results of this assessment to provide a series of recommendations on
the design of password recovery systems. Section \ref{Concluding remarks}
concludes the paper, including a discussion of possible directions for further
research. 	
\section{Prior Art } \label{Prior Art}

Many authors have looked at web password recovery, mostly focussing on
particular classes of recovery system or the shortcomings of widely used
approaches. 	
\begin{itemize}

\item Several authors have looked at ways to securely store backup copies
    of passwords on a client device; this is outside the scope of password
    recovery as we define it, since the web service itself is not involved.
    We briefly mention two papers of this type. Ellison et al.\
    \cite{ellisonprotectingsecret} describe how a locally stored copy of a
    password can be held in encrypted form, protected using answers to
    personal questions; the encryption key is derived from these questions
    in such a way that correct answers to \textit{k} of \textit{n}
    questions enables it to be reconstructed, allowing the user to forget
    some of the answers. Somewhat analogously, Frykholm and Juels
    \cite{frykholm2001error} propose a provably secure technique for
    fault-tolerant password recovery; a secret password is stored protected
    by a collection of low entropy secrets, such that recovery is possible
    with only a subset of these secrets. This again enables the user to
    maintain secured backup copies of passwords, and hence, like the
    Ellison et al.\ scheme, it is outside the scope of this paper.

\item  Chmielewski et al.\ \cite{chmielewski2009client} focus on what they
    call client-server password recovery. They propose a series of
    protocols that allow a user to automatically recover a password from a
    server using partial knowledge of the password, and prove their
    security in a formal model.  This can be regarded as a contribution to
    theoretical cryptography, rather than as a practical solution to the
    everyday problem of web password recovery.

\item Mannan et al.\ \cite{MannanBBLO11} propose a scheme for password
    recovery rather different from many commonly used approaches.  They
    propose that websites maintain copies of user passwords encrypted under
    a public key for which the user holds the private key on
    a personal mobile device (PMD).  When the user invokes the recovery
    service, the website sends the encrypted copy to the user, who uses his/her her PMD to decrypt it.
		
\item Kharudin et al.\ \cite{din2015password} describe a graphical user
    authentication method and propose its use for password recovery.
    However, the main focus is on the authentication technique rather than
    on the password recovery process.

\end{itemize}

All the prior art we have so far described, whilst relating to password
recovery, falls outside our scope here, since the proposals are either
independent of the web server or require major changes in how clients and
servers interact. However, as we describe next, some authors have addressed the
problem we consider here.

\begin{itemize}

\item A recent study by Stavova et al.\ \cite{stavova2016codes} examines
    the usability of two password recovery techniques, namely backup codes
    (primarily used as a recovery process for two-factor authentication)
    and the use of trusted associates (social authentication). They
    examined a particular scheme where the backup code is stored as a QR
    code, to address the issue of backup codes being forgotten and/or
    written in clear text. They also considered a case where the account
    holder has to call a client centre to get the first half of the
    password and a trusted associate has to call the client center to get
    the second half of the password.

\item Guri et al.\ \cite{guri2016personal} point out how details revealed
    during password recovery can be used to learn potentially sensitive
    personal user information.

\item A number of authors have examined the security of challenge
    questions, a secondary means of user authentication widely used in
    real-world password recovery (see section \ref{information}). In 2008,
    Rabkin \cite{Rabkin08} examined leaks of answers to these questions via
    social media. In the following year, Just et al.\
    \cite{just2009personal} performed extensive surveys of user behaviour
    to analyse the relative security offered by various questions. In
    parallel work, also published in 2009, Schechter et al.\
    \cite{schechter2009s} looked at the security properties of security
    questions as a secondary means of user authentication in the context of
    password recovery. In 2015, Bonneau et al.\ \cite{bonneau2015secrets}
    conducted a study using a real-world data set to examine the security
    and memorability of security questions. A yet more recent study by
    Gelernter et al.\ \cite{GelernterKMP17} describes a possible man-in
    the-middle (MitM) attack on password recovery by using the registration
    process for a malicious MitM website to gather the information needed
    (e.g.\ answers to personal questions) to conduct recovery for a
    different website.
	\end{itemize}
	
\section{A General Model} \label{A General Model}

We now present a general model for a password recovery system. It provides a
framework within which we can examine the security of various options for the
recovery process. As discussed above, password recovery is a process which
enables users to re-establish a password for a website account. Sometimes
referred to as \textit{password reset}\footnote{We avoid this terminology since
it implies changing the existing password, and not all password recovery
schemes involve such a change.}, password recovery is typically performed when
a user forgets the password for a website. There are a variety of ways in which
password recovery can be performed, some more secure than others, and the
general model is intended to capture all the means of password recovery in
current use.

\subsection{Constituent Processes}

A password recovery process typically involves three sub-processes, namely:

\begin{itemize}
\item \textbf{registration}, in which user-specific information to be used
    during password recovery is captured by the website,
\item \textbf{password setup}, in which the user chooses and sends the
    website a new password for that site, and
\item \textbf{recovery}, where the user interacts with the website with the
    goal of re-establishing a shared secret password.
\end{itemize}

Typically, the first of these will be performed just once, the second will
be performed infrequently, and the third will be performed whenever necessary.
These three sub-processes are next examined in greater detail. 		

\subsection{Registration}

In establishing an account with the website, a user typically sets up a user
name and password, as well as other information for use in the recovery
process. This information is used to ensure the security of the recovery
process, i.e.\ to help prevent an unauthorized party from changing the password
to cause a denial of service, or, even worse, obtaining a valid password for
the account. Existing categories of information of this type are examined in
section \ref{information}.

\subsection{Password Setup}

We assume the website uses a password to authenticate the user, and so both
initially (probably as part of registration) and whenever the user wishes to
change it, the user will need to supply the website with a password. Other
information, e.g.\ a `password hint', may be collected at the same time.
Further examples of information that might be gathered in this stage are given
in section \ref{password_setup}.

\subsection{Recovery}

Password recovery is typically invoked by a user when he/she forgets the
password for a website. Such a process typically has three stages, as follows.

\begin{enumerate}

\item \textbf{Recovery request}.  This involves the user signalling to the
    website to request password recovery. The website will typically
    provide a means for this to occur, e.g.\ a special button.
		
\item \textbf{Request validation}. The website checks that the request is
    valid. A website may perform user authentication at this stage,
    obviously using a means other than the password, although by no means
    all websites do this. The website may also attempt to verify that the
    request originates from a human rather than a bot, e.g.\ using a
    CAPTCHA \cite{AhnBHL03}.
		
\item \textbf{Password re-establishment}. At the conclusion of this stage,
    if successful, the user is equipped with a valid password for the
    website. There are two main implementation approaches.
    \begin{itemize}
    \item The password recovery system can help the user remember
        (recover) his/her password; this will typically involve the
        system keeping a copy of every user password, and password
        recovery will involve reminding the user what it is (typically
        after he/she has been authenticated).
    \item The password recovery system can help the user set up a new
        password, often referred to as \textit{password reset}. This
        may involve the system giving the user a temporary password,
        which the user must change at first use.
    \end{itemize}
\end{enumerate}

Regardless of the particular implementation approach, the recovery process will
typically involve using a secure communications channel to the user, e.g.\
based on a previously registered email address or phone number.  Examples of
how each of these steps might be executed are provided in sections
\ref{Recovery Request and Validation} and \ref{Password Re-establishment}.

\section{Model Components} \label{Model Components}

We next examine how the model components can be instantiated. The various
options introduced here are critically analyzed later in the paper.

\subsection{Registration} \label{information}

Registration for the password recovery is typically implemented as part of a
general registration process, in which the user establishes a new account
(known as \textit{registering} or \textit{signing up}). As well as gathering
information, it may also involve security-related steps, e.g.\ solving a
CAPTCHA to prevent automated user account harvesting. It often involves
collecting a wide range of information, including matters not relevant to
password recovery (e.g.\ payment information); we focus only on information
related to password recovery. Such information can be divided into two main
categories:

\begin{itemize}

\item \emph{personal information}, i.e.\ information about the individual
    that may be used for a range of purposes apart from password recovery;

\item \emph{recovery information}, i.e.\ information used only for password
    recovery.

\end{itemize}

A website will also typically require the user to choose a unique user name (or
give an email address to function as a user name) as well as a password; these
two pieces of information are obviously key to password recovery, although
given their special status we do not include them in this classification.  We
next look at examples of widely used information types of both categories.

\subsubsection{Personal information } \label{Example of Security Question}

The following are examples of the types of personal information that may be
gathered during registration and subsequently used for password recovery. In
each case, examples of websites are given which use the type of information
specified in a password recovery process.

\begin{itemize}

\item 	name (first name, last name), e.g.\ as gathered by
    Instagram\footnote{https://goo.gl/sG4wDv, accessed: 14/01/2018};

\item gender, e.g.\ as collected by
    Facebook\footnote{https://goo.gl/wCsLMk, accessed: 14/01/2018};

\item birth date (day, month, year), e.g.\ as gathered by Google
    Mail\footnote{https://goo.gl/b67119, accessed:14/01/2018};

\item street address (street name, city, country), e.g.\ as used by
    Microsoft email\footnote{https://goo.gl/9hbHNX, accessed: 14/01/2018};

\item email address, e.g.\ as collected by
    Amazon\footnote{https://goo.gl/yrv1fA, accessed: 14/01/2018};

\item	phone number (e.g.\ mobile number), e.g.\ as gathered by
    Twitter\footnote{https://goo.gl/uUYysF, accessed: 14/01/2018}.

\end{itemize}

\subsubsection{Recovery information} \label{4.1.2}

The types of information established purely for password recovery purposes vary
widely, depending on the detailed operation of the recovery process. We can
identify the following general categories.

\begin{itemize}

\item 	\textbf{Recovery authentication information}, i.e.\ information
    that can be used to authenticate the user. Examples include the
    following.

    \begin{itemize}
	\item \textbf{Answers to security questions}: the website may give
a list of security questions (also known as \textit{personal knowledge}
or \textit{challenge} questions), for a user-selectable subset of which
the user must provide answers. These questions cover topics which the
user can easily remember since they relate to the user's personal life,
e.g.\ school name, mother's maiden name, pet name, memorable street
address, birthplace, favorite colour, etc. Examples of websites that
request answers to security questions during registration include Apple
and Google. In addition, some websites, e.g.\ Alipay.com, allow users
to customize the question, i.e.\ the user provides both the question
and the answer.
\item \textbf{One-time recovery (backup) codes}: a website may set up
    one or more one-time backup codes, which the user must securely
    retain for possible future use for account recovery. These backup
    codes are typically not used for password recovery as we define it
    here, but for closely related purposes.  In particular, Facebook
    and Google both support the establishment of such codes to enable
    users to recover account access if a second authentication factor
    fails or is unavailable, e.g.\ if a one-time password sent via SMS
    or email cannot be accessed by the user.  Of course, such backup
    codes could also be used for password recovery but, since we are
    not aware of any websites using them in this way, we do not discuss
    them further in this paper.
    \end{itemize}

\item \textbf{Recovery contact details}, i.e.\ special contact details,
    such as an email address or phone number, that are used for password
    recovery. Examples of such details include the following.
    \begin{itemize}
    \item \textbf{Contact details for trusted associates (trustees)}:
        i.e.\ email addresses or phone numbers for one or more
        individuals trusted by the account holder, e.g.\ as used by
        Facebook. During password recovery, a verification code is sent
        to a trustee, who is trusted to relay it to the correct user.
    \item \textbf{Recovery email address}: i.e.\ one or more email
        addresses to which a one-time password or a link to a special
        recovery page is sent by the website during the recovery
        process.
    \item \textbf{Recovery phone number}: i.e.\ a phone number used to
        send a one-time password for recovery purposes.
	\end{itemize}

\item \textbf{Recovery preferences} can be gathered at this stage if a
    website offers more than one option for password recovery.  The user
    preference can be established during registration, or, as is the case
    for Google, the user is simply offered various options if the password
    recovery process is invoked.
	\end{itemize}

\subsection{Password Setup}	\label{password_setup}

As part of the process of entering a password, the website may gather and store
the following types of information.

\begin{itemize}

\item \textbf{Password hints}: when a password is entered, some websites
    also request a hint, intended to help the user remember the password.
    Alternatively, some websites, e.g.\ Google (see figure \ref{Google
    password change hint}), record the time when the password was last
    changed, and subsequently use this information as a password hint when
    the user makes a password recovery request.

\item \textbf{Old passwords}: some websites, e.g.\ Gmail, keep old
    (superseded) passwords, to enable their use for authentication during
    password recovery.

\end{itemize}

\begin{figure}[H]
	\begin{center}
		\includegraphics[width=2.8in]{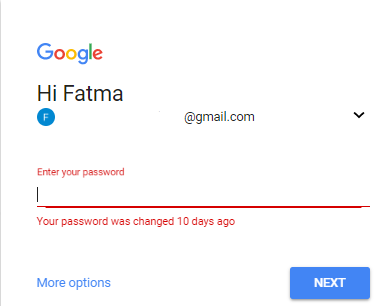}
		\caption{Google password change hint}
		\label{Google password change hint}
	\end{center}
	\end{figure}
	
\subsection{Recovery Request and Validation} \label{Recovery Request and Validation}

We consider the request and validation steps together since, in practice, these
steps are often combined. A website will typically provide a simple means for a
user to invoke the password recovery process, e.g.\ by providing a link
somewhere close to the password field in the login page. Once invoked the user
will typically be asked to perform one or more of the following steps in order
to prevent the acceptance of fraudulent requests:

\begin{itemize}

\item solve a CAPTCHA --- some websites, e.g.\ PayPal, use a CAPTCHA during
    password recovery to filter out automated attacks;
		
\item answer one or more of the pre-established security questions, e.g.\
    as performed by Apple;

\item select an option for password recovery (if the user registered
    multiple methods), e.g.\ as is the case for Amazon;

\item enter the last password the user can recall, e.g.\ as requested by
    Google.

\end{itemize}

Following initial acceptance of the request, a user may be required to engage
in further interactions to validate the request, e.g.\ as follows.

\begin{itemize}

\item An email is sent to the user's registered address, and the user is
    asked to perform a task using information contained in the email, e.g.\
    clicking on an embedded link or entering a code value.

\item An SMS (text) message containing a secret code is sent to the user's
    registered mobile number, and the user is requested to enter this code;
    alternatively the user may receive an automated call to his/her mobile
    phone and be asked to engage in a short dialogue.

\end{itemize}

\subsection{Password Re-establishment} \label{Password Re-establishment}

We next review some commonly used methods for re-establishing a password.

\begin{itemize}
\item \textbf{Email reset} involves a website sending a message containing
    secret information to the registered email address of the user
    requesting recovery.  Subsequent use of this information implicitly
    authenticates the user, assuming that only the user can access emails
    sent to his/her email address. This approach is very widely used;
    Bonneau et al.\ \cite{BonneauP10} found that 92\% of the 138 websites
    they tested use email-based password recovery. Various types of secret
    information are commonly sent, e.g.\ as follows.
	\begin{itemize}

    \item \textbf{Verification codes/temporary passwords} are a
        temporary means of accessing the user account, purely for the
        purposes of establishing a new password, and are used, for
        example, by Amazon and Wikipedia --- see Figure \ref{Wikipedia
        password recovery using temporary password}.  Note that
        Wikipedia limits users to one temporary password per 24-hour
        period, and its temporary passwords expire after a week.

\begin{figure}[htb]
				\begin{center}
					\includegraphics[width=2.8in]{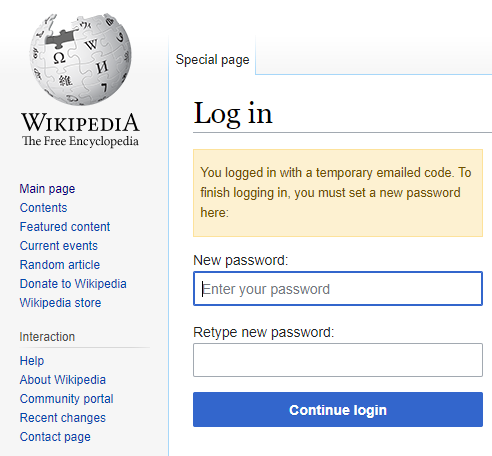}
					\caption{Wikipedia password recovery using temporary password}
					\label{Wikipedia password recovery using temporary password}
				\end{center}
			\end{figure}

\item \textbf{Links} (URLs) are embedded in emails sent to a
    user-registered address. Clicking on such a link (which typically
    contains a secret string) redirects the user browser to a page
    enabling the user to set up a new password. Some websites, e.g.\
    Twitter, limit the validity period of such links.

\end{itemize}

\item \textbf{SMS reset} messages are sent to the user's registered mobile
    number, and typically contain a secret verification code (analogously
    to email reset).

\item \textbf{Use of an old password} is permitted by some websites, e.g.\
    Google, as a means of authenticating a user for password recovery.
    Typically, it will form only part of the process of authenticating the
    user.

\item \textbf{Use of a trustee} is supported by some websites (e.g.\
    Facebook), where the trustee is used as a secure communications channel
    to the account holder channel for sending secret recovery information,
    such as a temporary (one-time) password.

\end{itemize}
	
\subsection{After Password Re-establishment}

After a user has reset his/her password, some websites immediately log out
the user and require a fresh log-in with the new password; others allow the
user to continue without logging-in again. In parallel with this, some websites
notify the account holder via email that password recovery has occurred, with
the goal of alerting users if the recovery was not authorised by the account
holder.

\section{Security and Usability Issues} \label{Security and usability issues}

We now consider the security and usability issues for the recovery validation and password re-establishment techniques given in sections
\ref{Recovery Request and Validation} and  \ref{Password Re-establishment}.

\subsection{Security Questions} \label{security Question}

As observed in section \ref{Recovery Request and Validation}, security
questions are widely used to help ensure that a recovery request is valid. As
discussed in section \ref{Prior Art}, apart from the fact that gathering such
personal data potentially endangers user privacy, a range of serious security
and usability issues have been identified. Bonneau et al.\
\cite{bonneau2015secrets} identify the following problems.

\begin{itemize}
\item  Answers to some questions are more readily guessed than others since
    they have a small answer space; such questions offer relatively little
    protection.
\item Some users may provide false answers with the goals of limiting what
    is revealed about them and making it harder for impostors to guess
    their responses. However, users may forget their false answers, making
    the security questions useless in the recovery process.

\item Although the questions are designed to cover topics which users will
    always know, some questions may nevertheless require the user to
    remember the answer they gave, e.g.\ which of many pet names they
    chose, or which colour they said was their favourite. Given such
    questions are likely to be used very infrequently, and considerable
    time may elapse between setting them up and using them, a user's
    recollection of the `correct' answers is likely to fade.

\item Cultural differences can have an impact on the memorability of
    different type of questions, e.g.\ Bonneau et al.\
    \cite{bonneau2015secrets} found that, for a typical set of questions,
    French users are most likely to recall their first phone number and are
    least likely to recall their father's middle name.

\end{itemize}

Also, the answers to some questions could be obtained via social media, as
discussed by Rabkin \cite{Rabkin08}. Even more seriously, the same questions
are used by many sites, making possible the type of MitM attack described by
Gelernter et al.\ \cite{GelernterKMP17} (see section \ref{SMS Reset}), where a
malicious site persuades a user to register with answers to security questions
which are then used to impersonate the user to another site. The real issue
here is that the authentication information (answers to security questions) is
not site-specific; analogous problems arise when users employ the same password
with many sites --- this has caused many well-documented security problems,
notably when passwords for one site are compromised and can then be used to try
to impersonate users to other sites\footnote{This analysis suggests a very
simple attack on passwords, where a malicious entity sets up a site and
persuades users to register and choose an ID and password; the site can then
act on the assumption that some users will employ the same user name/password
combinations elsewhere, and can try them out with other sites to see if they
work. Such an attack could be very effective without even requiring any
real-time MitM activity or compromise of existing password databases.}. 	

The main lessons from this analysis would appear to be that: (a) security
questions should be carefully chosen to maximise the level of security offered;
(b) security questions only offer a limited level of security and should
always be used in combination with other methods; and (c) ideally
authentication information should somehow be made site-specific. 	

\subsection{Trustee-based Recovery}

As discussed by Gong and Wang \cite{GongW14}, there are a number of security
issues with relying on trusted `friends' of a user to support password
recovery.

\begin{itemize}
\item The account holder may forget who they chose as trustees, so that an
    individual may remain a nominated trustee even if they are no longer
    trusted.  That is, the state of trust can change with time and a user
    could forget to remove a trustee from list.
\item Use of trustees can be prone to `forest fire attacks' \cite{GongW14}, where
    compromise of a trustee account can compromise many other accounts.
\item Users are sometimes obliged to nominate several trustees, e.g.\
    Facebook mandates three.  This may force the user to select less
    trusted associates.
\item A malicious trustee could take over an account by triggering password
    recovery and (if necessary) comunicating with other trustees (e.g.\
    using social engineering) to obtain all the secret information
    necessary.
\end{itemize}

\subsection{Email Reset} \label{Email Reset}

A major vulnerability of this approach is its assumption that emails cannot be
intercepted. There are various ways this assumption could be invalidated.

\begin{itemize}
\item If emails are retrieved via an unencrypted link, e.g.\ if the user
    accesses email via a browser and the website does not use https, or if
    a mail client employs IMAP over a link not protected using SSL/TLS,
    then emails might be intercepted.  This threat is particularly
    significant when using public access networks; a MitM attacker
    operating a `fake' public wireless access network could intercept
    emails received via SMTP.
\item If the reset email is forwarded to a third party, either accidentally
    or deliberately, then the contents could be compromised.
\item Malware in the client device could be used to obtain the reset email.
    For example, Gelernter et al.\ \cite{GelernterKMP17} describe a
    malicious android application which, if installed, can covertly read
    received emails.
\end{itemize}

		The use of email for distributing reset information also has other
risks. For example, consider the case where the website for which password
recovery is being performed is itself a provider of an email service (e.g.\
Google). In this case, sending a password reset message by email will not help,
as the user will not be able to log in to retrieve it.  In such a case it is
common practice for the user to be asked to register an alternative email
address to be used to distribute recovery information.  During password
recovery, in some cases the user will be shown a partially obfuscated version
of this alternative email address, and the user will be asked to confirm that
this is the correct address for use in password recovery.  Figures \ref{Partial
leakage of user contact number} and \ref{Partial leakage of alternative email
address} show this process as used by Gmail.  This is a clear privacy breach,
in that it may well be possible for a third party to learn the entire
alternative address, e.g.\ by a web search or by automated harvesting of email
addresses \cite{guri2016personal} (see, for example, Polakis et al.\
\cite{PolakisKAGPM10}).

Apart from the risk of spam, restricting access to email addresses is clearly
desirable given that email addresses are often used as user identifiers. A
further possible problem arising with the use of email relates to the
non-permanent nature of email addresses.  A user may register an email address
which may later become re-assigned to someone else, e.g.\ because they change
service providers or employers.  In such a case, a password reset message may
be received by a third party, who could use it to take control of the user
account. Another issue could arise in a work environment in which employee
emails are temporarily forwarded to another person, e.g.\ because the employee
is sick or on leave; this could be especially hazardous if the forwardee is
temporary or a contract worker.
	\begin{figure}[!tbp]
		\centering
		\begin{minipage}[b]{0.4\textwidth}
			\includegraphics[width=\textwidth]{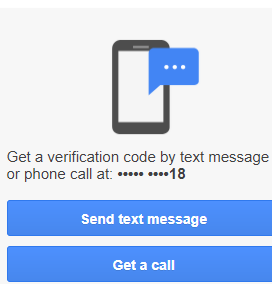}
			\caption{Partial leakage of user contact number}
			\label{Partial leakage of user contact number}
		\end{minipage}
		\hfill
		\begin{minipage}[b]{0.4\textwidth}
			\includegraphics[width=\textwidth]{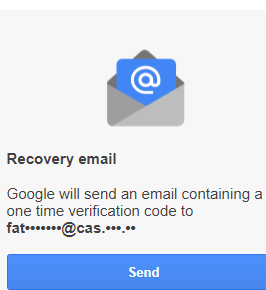}
			\caption{Partial leakage of alternative email address}
			\label{Partial leakage of alternative email address}
		\end{minipage}
	\end{figure}
	
\subsection{Verification Codes and Reset Links}

As discussed in section \ref{Password Re-establishment}, there are two main
ways email reset can be used for password recovery, namely by sending a
verification code or a reset link.  As discussed above, both give rise to a
risk if the reset email is compromised. Apart from the direct compromise
threat, both have other security and usability issues.

\begin{itemize}
\item	The following issues are associated with verification codes.
\begin{itemize}
\item Lack of entropy: since verification codes are typically short
    numeric values, they are potentially vulnerable to brute force
    searching attacks.  That is, if a website does not limit the number
    of attempts to enter the code, then a simple brute force attack
    becomes possible, as demonstrated by a successful attack on two
    Facebook sites \cite{Telefacebook}.
\item As discussed in greater detail in section \ref{SMS Reset} in the
    context of codes sent via SMS, a user could be misled into
    revealing a verification code to a malicious site in a type of
    phishing attack.
\end{itemize}

\item Reset links have a slightly different set of associated issues.
\begin{itemize}
\item Some websites do not expire unused reset links, i.e.\ they remain
    valid indefinitely.  For example, in an experiment we discovered
    that a link for rosegal.com remained valid for at least five
    months. Since it is hard to understand why a link needs to be valid
    for more than a few hours, this represents an unnecessary risk,
    since if such a link is ever disclosed it may enable the associated
    account to be hijacked.
\item Karlof et al.\ \cite{KarlofTW09} describe a phishing attack on
    reset link-based password recovery.  The attacker starts password
    recovery on a website for a target user, knowing this will cause
    the website to send an email containing a link to the target user.
    Simultaneously the attacker emails the user asking them to paste
    the link in an email they are about to receive into an
    attacker-controlled website; a similar attack works against
    verification codes in emails. This attack is similar to the
    Gelernter et al.\ \cite{GelernterKMP17} attack on SMS-based
    password recovery (see section \ref{SMS Reset}).
\end{itemize}

\end{itemize}

\subsection{SMS Reset} \label{SMS Reset}

There are a variety of issues associated with the use of SMS reset messages,
many of which are analogous to issues associated with email reset.

\begin{itemize}
\item Since codes sent to mobile phones are typically only six numeric
    digits\footnote{Google and Dropbox both use 6--digit verification
    codes}, i.e.\ there are only $10^{6}$ possibilities, this means that it
    may be possible to successfully guess a code value.  It is therefore
    imperative that websites limit the number of entry attempts a user is
    permitted.
\item As noted in section \ref{security Question}, Gelernter et al.\
    \cite{GelernterKMP17} describe an MitM attack on password recovery: the
    registration process for a malicious website is used to gather the
    answers to personal security questions in order to successfully
    complete request validation for a website with which the user already
    has an account. Of course, this MitM step does not complete the attack,
    as the attacked website will typically, after receiving the correct
    answers to security questions, send a verification code (e.g.\ via SMS)
    to the genuine user. To complete the attack, the malicious website (as
    part of its registration procedure) tells the user to expect a message
    containing a verification code which should be entered into the
    registration page. If the user enters the verification code received
    from the attacked website, the malicious website will now have what it
    needs to complete the process of capturing the user's account. The only
    defence against this attack is for the message containing the
    verification code to make it clear which website it is intended for
    (i.e.\ the attacked website rather than the malicious website).

    This attack will work against verification codes sent via email or SMS,
although it is much harder for the sending website to make clear how the
code should be used in a 160-character SMS message than in an email.
Indeed, Gelernter et al.\ \cite{GelernterKMP17} describe the results of a
detailed survey of the use of SMS-based verification codes, with the goal
of understanding both how clear the SMS messages are, and how likely users
are to mistakenly enter a code intended for one website into a box on a
different website. They go on to make a series of recommendations regarding
how SMS messages should be designed to minimise the likelihood that a user
can be deceived.  They also recommend that the veriirefication code should
have a short validity period.

\item In some countries, network coverage in rural areas is often poor or
    non-existent, meaning it may not be possible for a user to obtain a
    verification code sent via SMS. Similarly, if a user is traveling,
    he/she may not be able to receive the SMS. Also, occasionally SMS
    messages are delayed, which may cause the user to initiate many
    requests.

\item Just as for email reset, the security of SMS reset relies on the
    assumption that an SMS message is a secure channel to the user.  Again,
    as for email reset, this assumption can be invalidated in a range of
    ways, e.g.\ as follows.
		\begin{itemize}
		\item 	Guri et al.\ \cite{GelernterKMP17} describe an attack
based on an apparently harmless Android app which can monitor and
redirect SMS messages to an attacker server.  If such an app was
installed, then a password recovery SMS might be compromised.
		\item 	If the phone to which the SMS reset is sent is stolen
then the contents of the SMS reset message could be compromised; even
if the phone is password-locked, the message may still be compromised
since many phones display the contents of received SMS messages without
requiring the phone to be unlocked.
		\item 	SMS reset messages could be intercepted at the network
operator, e.g.\ by a malicious employee, or when sent over the air
interface if the air interface link is not encrypted (as is the case in
some countries). The reset message could also be intercepted by an
unauthorised base station (an IMSI catcher or `Stingray')
\cite{Lilly17} or by exploiting  weaknesses in implementations of the
SS7 protocol used by telecomunication companies to communicate with
each other \cite{Welch17}.
		\item If the user changes his/her phone number and forgets to
notify the website, then the SMS reset message will be received by the
new owner of that number.
		\end{itemize}
\end{itemize}

\subsection{Other Issues}

We conclude this discussion of existing password recovery techniques by briefly
reviewing certain other issues which have a bearing on the security and/or
usability of the password recovery process.

\begin{itemize}
\item Use of an old password as the only means of user authentication for
    password recovery is clearly dangerous, since one reason a user might
    replace a password is because it has been, or it is suspected of having
    been, compromised.  It also seems relatively unlikely that a user will
    recollect an old password if they cannot remember their current
    password.
\item Many web services use an email address as a user ID, e.g.\ Amazon,
    Facebook, Dropbox, Instagram, Twitter and LinkedIn.  This has the major
    advantage of making it easy for the user to remember their account
    identifier, and it is also convenient for web service providers to
    deliver account-related functions via email as well as simplifying
    registration.  However, as discussed by Jin et al.\ \cite{JinTJ10},
    there are significant risks associated with such an approach, not least
    that an attacker automatically knows one of the two parts of a user
    credential set, thereby facilitating attacks on the password recovery
    process.

\item Password hints may reveal sensitive information, potentially
    endangering user security or privacy. For example, the Google password
    hint shown in figure \ref{Google password change hint} indicates when
    the user last changed his/her password -- this reveals that Google
    was being used at that time. Whilst this may not seem so significant,
    for some sites this could be far more revealing.

\end{itemize}

\section{Towards Secure Password Recovery} \label{Towards secure password recovery}

We next provide recommendations on best practice in implementing a password
recovery system.  These recommendations are based on the analysis given in the
previous section.  We consider the two most security-significant steps in
password recovery, namely request validation and password re-establishment.

\subsection{Recovery Request Validation}

As discussed in section \ref{Recovery Request and Validation}, two main
security-related steps are commonly used during request validation, namely use
of a CAPTCHA and security questions.  The CAPTCHA prevents automated attacks on
password recovery, and appears a reasonable step to include.  Security
questions are used to prevent malicious triggering of password recovery, which
could otherwise be done on a large scale, e.g.\ using databases of email
addresses (since user names are often the same as email addresses).  However,
the use and effectiveness of security questions is questionable for a variety
of reasons. Firstly, as discussed in section \ref{security Question}, the
answers can sometimes be readily guessed or obtained via social media, and,
most seriously, the MitM attacks of Gelernter et al.\ \cite{GelernterKMP17}
suggest that they offer relatively little protection.  Secondly, there are
significant privacy issues, since use of such questions involves gathering
personal data, which could be misused.  Thirdly, there are also usability
issues, in that not all answers to questions can always be readily recalled. It
is therefore highly questionable whether using security questions as a
filtering mechanism is worth the trouble, since while it offers limited
security it also has significant negative privacy and usability disadvantages.

\subsection{Password Re-establishment}

Depending on how password re-establishment works, there are a number of
important recommendations which emerge from the analysis in section
\ref{Security and usability issues}.  We look separately at three main areas:
email reset, SMS reset, and the use of trustees.

\subsubsection{Email reset}

If email reset is used, then the website must try to minimise the chance that
the email reaches an incorrect recipient.  This means, for example, that the
email address to which the message is to be sent should ideally have been used
(or confirmed by the user) recently.  This could be achieved by asking the user
to enter the email address to which the reset email should be sent, and only if
this matches with the address stored by the website should it be used.

If verification codes are sent via email, then the discussion in section
\ref{Email Reset} suggests that: (a) they should use as large an alphabet as
possible (e.g.\ including letters, digits and punctuation), and be as long as
possible, subject to usability constraints; and (b) they should expire after a
short period of time.

Reset URLs sent in an email should: (a) have a short expiry period (just as for
codes); (b) contain a random (secret) string which should be sufficiently long
to make successful guessing attacks highly improbable; and (c) start with
`https', so that the connection to the server is secure.

\subsubsection{SMS reset}

If SMS is used to deliver the means to re-establish a password, then very
similar requirements to those for email reset apply.  That is, regardless of
what information is sent in the SMS, the website should try to ensure that the
phone number is current, e.g.\ by requesting the user to re-enter the number,
and only if it matches with the number stored by the website should it be used.
Apart from the requirements applying to verification codes and links previously
mentioned, Gelernter et al.\ \cite{GelernterKMP17} give a list of
recommendations aimed specifically at verification codes sent via SMS.  In
particular: (a) the SMS message should be designed so that the code is not
displayed on the phone screen when locked, i.e.\ forcing the user to unlock the
phone to obtain the code; and (b) the wording of the SMS should minimise the
chance that the user can be deceived into submitting the code to the wrong
website, including indicating the identity of the sending website, the purpose
of the message, and a warning not to divulge the code to any other person or
website.

\subsubsection{Trustee-based password recovery}

Since significant risks arise from out of date trustee data, as part of the
recovery request process the account holder should be asked to verify contact
details for the trustee(s) to be used. For example, the user could be asked to
re-enter trustee email addresses or phone numbers, with the website only using
them if they match already stored values. Web sites could also periodically
require users to revalidate trustee data. Clear instructions should be sent
with the password recovery information sent to a trustee, to minimise the
chance that a trustee is misled (e.g.\ via social engineering) into disclosing
the recovery data to the wrong party.

\section{Concluding Remarks} \label{Concluding remarks}


We have given a general model for the password recovery process, and we have
also examined the range of ways in which this model is instantiated by today's
websites.  We then provided the first comprehensive review of known security
and privacy weaknesses in existing approaches to password recovery; we also
examined usability issues.  This then allowed us to make a series of
recommendations regarding how best a password recovery process should be
designed.


There is clearly a need for further research in this important area, as well as
new, more secure ways, of performing recovery. Almost all the techniques in
common use are to some extent flawed, and many also pose privacy risks. One
direction for future work would be to conduct large scale practical trials to
try to understand better how users can interact with password recovery systems
both securely and reliably. A further issue which has hardly been explored is
that of account recovery, i.e.\ where a user forgets both his/her user name and
password, or a second authentication factor fails.  A range of techniques are
in use for this purpose, e.g.\ backup codes (see \ref{4.1.2}), and the
usability and security of these techniques clearly merits further
investigation.

\bibliographystyle{plain}
\bibliography{PR}

\end{document}